\documentclass[aps,preprint]{revtex4}
\usepackage{amsfonts}
\usepackage{mathrsfs}
\usepackage{amsmath}
\usepackage{amssymb}
\usepackage{subfigure}
\usepackage{graphicx}
\usepackage{subfigure}
\usepackage{epstopdf}
\usepackage{hyperref}
\usepackage{float}
\begin{document}
\title{Thermodynamics and phase transition in central charge criticality of charged Gauss-Bonnet AdS black holes}

\author{Yang Qu}
\email{quyang@stu.scu.edu.cn}
\author{Jun Tao}
\email{taojun@scu.edu.cn}
\author{Huan Yang}
\email{yanghuan2@stu.scu.edu.cn}

\affiliation{Center for Theoretical Physics, College of Physics, Sichuan University, Chengdu, 610065, China}

\begin{abstract}
In this paper, we investigate the thermodynamics of D-dimensional charged Gauss-Bonnet black holes in anti-de Sitter spacetime. Varying the cosmological constant, Newton constant and Gauss-Bonnet coupling constant in the bulk, one can rewrite the first law of thermodynamics for black holes. Furthermore, we introduce the central charge and study the critical behaviors, which show the apparent discrepancy from other black holes. Based on this approach, we disclose the phase transition structures in $D=4, 5$, and $6$. Besides, a triple point where the small/intermediate/large black holes can coexist is found in $D=6$.
\end{abstract}
\maketitle

\section{Introduction}
General relativity is the most powerful theory to describe spacetime. The study of black holes and their thermodynamics helps us understand the nature of gravity. Black holes in asymptotically anti–de Sitter (AdS) spacetime demonstrate the thermodynamic phase transition behavior in thermal equilibrium such as Hawking-Page phase transition \cite{Hawking:1982dh}, and then the black holes thermodynamics with phase transition have been extended to a variety of more complicated backgrounds, such as charged AdS black holes spacetime \cite{Chamblin:1999hg,Chamblin:1999tk}. 

The recent development of black holes thermodynamics reveals that the negative cosmological constant can be treated as a variable and related to the thermodynamic pressure. In this extended phase space, the new thermodynamic behavior and phase transition structures of AdS black holes were discovered \cite{Kastor:2009wy,Dolan:2012jh,Kubiznak:2012wp,Gunasekaran:2012dq,Mo:2018rgq,Anabalon:2018qfv,Miao:2018fke}. However, there are some problems with the holographic interpretation of black holes thermodynamics \cite{Johnson:2014yja,Dolan:2014cja,Kastor:2014dra,Zhang:2014uoa,Zhang:2015ova,Dolan:2016jjc,McCarthy:2017amh}. For example, the first law of black holes thermodynamics cannot be straightforwardly related to thermodynamics of the holographic dual conformal field theory (CFT) \cite{Karch:2015rpa,Sinamuli:2017rhp,Visser:2021eqk}. The study of thermodynamics in AdS black holes by using the AdS/CFT correspondence can help us understand the dual conformal field theory at finite temperature \cite{Maldacena:1997re}. The variation of cosmological constant $\Lambda$ relates to both the variation of central charge $C$ and the CFT volume $\mathcal{V}$ of boundary CFT (the AdS radius corresponds to the boundary radius). It shows that the law of thermodynamics need to be modified when we consider the volume and its conjugate, i.e. the thermodynamic pressure in the bulk. Moreover, variations of the fundamental constants, such as Newton constant and the cosmological constant, will imply that more fundamental theories exist \cite{Gibbons:1996af, Creighton:1995au}. To deal with this problem, Cong et al first show that varying Newton constant $G$ along with the cosmological constant $\Lambda$ can keep the central charge $C$ of the dual CFT fixed \cite{Cong:2021fnf}. Therefore the first law can be written as the mixed form which includes the appropriate new thermodynamic volume and chemical potential. It establishes a connection between the holographic and bulk thermodynamics by using the duality relation of the central charge. The research on the free energy reveals the phase structures with a critical value of the central charge, above which the phase diagram has a swallowtail behavior that represents a first order phase transition. Following the same step, Newton constant $G$ can be regarded as a variable, so the central charge can be introduced into the first law and the thermodynamics were discussed for different black holes \cite{Alfaia:2021cnk,Zeyuan:2021uol,Gao:2021xtt,Sadeghi:2022jlz,Kumar:2022fyq,Wang:2022hzh,Lobo:2022eyr,Ghosh:2022chx,Bai:2022uyz,Dutta:2022wbh}. Then the phase transitions and criticality of dual CFT for charged AdS black holes were studied in phase space \cite{Cong:2021jgb,Rafiee:2021hyj}. 

The complete quantum gravity can be built up by studying the correction of higher order derivatives. The key of extending the general relativity is to find the tensor qualified with these properties: (1) It is symmetric; (2) It is divergence free; (3) There is no higher derivative of metric in EOM (ghost free). Lovelock \cite{Lovelock:1971yv} found a general form of tensor satisfying above properties, which depends on the dimension of spacetime, yielding
\begin{equation}
\begin{aligned}
\mathcal{L}&=\sum_{i}^{n}a_{i}\mathcal{L}_{i},
\end{aligned}
\end{equation}
where $a_i$ is constant coefficient, $\mathcal{L}_i$ is the Euler density, $\mathcal{L}_0=1$, $a_0$ is the cosmological constant, $\mathcal{L}_1$ is the Einstein Hilbert term and $\mathcal{L}_2=R_{abcd}R^{abcd}-4R_{ab}R^{ab}+R^{2}$ is the Gauss-Bonnet gravity. There are many researches on Gauss-Bonnet gravity \cite{Odintsov:2020sqy,Oikonomou:2020sij,Oikonomou:2021kql}. Then Gauss-Bonnet black holes solutions \cite{Banados:1993ur,Crisostomo:2000bb,Cai:1998vy,Aros:2000ij,Boulware:1985wk,Cai:2001dz,Wiltshire:1985us,Hendi:2017lgb} and their thermodynamic properties \cite{Cvetic:2001bk,Cai:2013qga,Cai:1998vy,Wei:2014hba,Wang:2019urm,Hendi:2016yof} were studied. The research of charged Gauss-Bonnet AdS black holes thermodynamics shows that the black holes phase transition exists in any dimension. 

For $D=6$, there exist the triple points and small/intermediate/large black hole phase transitions with the variation of the pressure $P$, for some parameters of charge $Q$. Generally we consider the Gauss-Bonnet term as a topological invariant in $D<5$, so no Gauss-Bonnet black holes exist in $D=4$. However, recently by rescaling the Gauss-Bonnet coupling constant $\alpha\rightarrow \frac{\alpha}{D-4}$, and taking the limit $D\rightarrow 4$, the four-dimensional non-trivial black holes solution can be obtained \cite{Glavan:2019inb}. Moreover, its thermodynamics in AdS spacetime have been studied \cite{Fernandes:2020rpa,Wei:2020poh}. Recently, the modified first law of thermodynamics for Gauss-Bonnet AdS black holes in $D\geq 5$ were studied \cite{Kumar:2022afq}, and they mainly showed that the universal property of critical charge $C$ was broken, while the phase transition structure has not been specifically studied. In this paper, we derive the modified first law of thermodynamics for charged Gauss-Bonnet AdS black holes and detailed study the properties of critical central charge and phase transition behaviors in $D\geqslant4$. Besides, we have discovered some interesting phase structures in $D=6$.

Our paper is organized as follows. In Section \uppercase\expandafter{\romannumeral2}, we give a brief review of thermodynamics for charged Gauss-Bonnet AdS black holes. In Section \uppercase\expandafter{\romannumeral3}, we derive the modified first law of thermodynamics for charged Gauss-Bonnet AdS black holes. In Section \uppercase\expandafter{\romannumeral4}, properties of critical central charge and phase transition structures in $D=4$ and $D=5$ are studied. In Section \uppercase\expandafter{\romannumeral5}, we find that the triple point and three-phase structures exist for some parameters for $D=6$. In section \uppercase\expandafter{\romannumeral6}, some discussions and conclusions are given.

\section{Thermodynamics of charged Gauss-Bonnet AdS black holes}
In this section, we will briefly review the thermodynamics of charged Gauss-Bonnet black holes in AdS space. The action of $D$-dimensional Einstein-Maxwell theory with the negative cosmological constant and Gauss-Bonnet term is
\begin{equation}
\label{the d-dimensional action}
\begin{aligned}
S&=\frac{1}{16\pi G}\int d^{D}x\sqrt{-g}\left[R-2\Lambda+\alpha(R_{\mu\nu\lambda\rho}R^{\mu\nu\lambda\rho}-4R_{\mu\nu}R^{\mu\nu}+R^{2})-4\pi  F^{2}\right],
\end{aligned}
\end{equation}
where $F_{\mu\nu}=\partial_{\mu}A_{\nu}-\partial_{\nu}A_{\mu}$ is the strength of the Maxwell field, $G$ is the cosmological constant, $\alpha$ is the Gauss-Bonnet coupling constant with dimension $(length)^2$ and positive in the string theory \cite{Boulware:1985wk}. The general static solution of this action is 
\begin{equation}
\label{line element of d Gauss-Bonnet black holes}
\begin{aligned}
ds^{2}&=-f(r)dt^{2}+\frac{1}{f(r)}dr^{2}+r^{2}h_{ij}dx^{i}dx^{j},
\end{aligned}
\end{equation}
where $h_{ij}dx^{i}dx^{j}$ is the line element of $(D-2)$-dimensional hypersurface with constant curvature $(D-2)(D-3)k$. The value of $k$ can be taken as $-1,0,1$ to represent the spherical, Ricci flat, and hyperbolic topology of the horizon of black holes, respectively. To simplify the analysis, we consider the spherical topology of the horizon, i.e. $k=1$, so the metric takes the form as
\begin{equation}
\label{metric sphere}
\begin{aligned}
ds^{2}&=-f(r)dt^{2}+\frac{1}{f(r)}dr^{2}+r^{2}(d\theta^{2}+sin^{2}\theta d\phi^{2}+cos^{2}\theta d\Omega_{D-4}^{2}).
\end{aligned}
\end{equation}

The metric function for charged Gauss-Bonnet AdS black holes is \cite{Boulware:1985wk,Cai:2001dz,Wiltshire:1985us,Cvetic:2001bk}
\begin{equation}
\label{metric function 1}
\begin{aligned}
f(r)&=1+\frac{r^{2}}{2\widetilde{\alpha}}(1-\sqrt{1+\frac{64\pi\widetilde{\alpha}GM}{(D-2)\Sigma_{k}r^{D-1}}-\frac{8\widetilde{\alpha}Q^{2}G}{(D-2)(D-3)r^{2D-4}}-\frac{4\widetilde{\alpha}}{l^{2}}}),
\end{aligned}
\end{equation}
where $\widetilde{\alpha}=(D-3)(D-4)\alpha$, $M$ is the mass of black holes, $Q$ is the charge of black holes, and for simplicity $\Sigma _k=1$ is the area of unit $(D-2)$-dimensional sphere .

Solving $f(r_+)=0$ to obtain the horizon radius $r_+$, we obtain the mass $M$
\begin{equation}
\label{mass1}
\begin{aligned}
M&=\frac{(D-2)r_{+}^{D-3}}{16\pi G}(1+\frac{\widetilde{\alpha}}{r_{+}^{2}}+\frac{r_{+}^{2}}{l^{2}})+\frac{Q^{2}}{8\pi(D-3)r_{+}^{D-3}},
\end{aligned}
\end{equation}
and the Hawking temperature $T$ can be expressed in terms of $r_+$ as
\begin{equation}
\label{temperature1}
\begin{aligned}
T&=\frac{\frac{(D-1)r_{+}^{4}}{l^{2}}+(D-3)r_{+}^{2}+(D-5)\widetilde{\alpha}-\frac{2Q^{2}G}{(D-2)r_{+}^{2D-8}}}{4\pi r_{+}(r_{+}^{2}+2\widetilde{\alpha})}.
\end{aligned}
\end{equation}

For thermodynamics of black holes, the expressions of the negative cosmological constant and its relative thermodynamic pressure are \cite{Kastor:2009wy}
\begin{equation}
\label{the pressure and the cosmological constant}
\begin{aligned}
\Lambda=-\frac{(D-1)(D-2)}{2l^{2}},\quad P=-\frac{\Lambda}{8\pi G},
\end{aligned}
\end{equation}
where $l$ is the radius of the $D$-dimensional AdS space. The mass of black holes could be considered as the enthalpy,  $M\equiv H$, rather than the internal energy of the  gravitational system \cite{Kastor:2009wy}. Moreover, the entropy of black holes takes the form as \cite{Cai:1998vy}
\begin{equation}
\label{entropy1}
\begin{aligned}
S&=\int_{0}^{r_{+}}T^{-1}(\frac{\partial M}{\partial r})_{Q,P}dr=\frac{r_{+}^{D-2}}{4G}(1+\frac{2(D-2)\widetilde{\alpha}}{(D-4)r_{+}^{2}}),
\end{aligned}
\end{equation}
and other thermodynamic quantities are
\begin{equation}
\label{potential and pressure}
\begin{aligned}
V=\frac{r_{+}^{D-1}}{D-1},\quad\Phi=\frac{Q}{4\pi(D-3)r_{+}^{D-3}}.
\end{aligned}
\end{equation}

The differential form of the first law of thermodynamics can be expressed as \cite{Cai:2013qga}
\begin{equation}
\label{D-dimensional first law}
\begin{aligned}
\delta M&=T\delta S+V\delta P+\Phi\delta Q+\mathcal{A}\delta\widetilde{\alpha},
\end{aligned}
\end{equation}
where $\mathcal{A}=(\frac{\partial M}{\partial\widetilde{\alpha}})_{S,P,Q}$ is the conjugate variable of $\widetilde{\alpha}$ in the extended space.
By the scaling argument, the smarr relation of black holes is \cite{Cai:2013qga}
\begin{equation}
\label{D smarr relation}
\begin{aligned}
(D-3)M&=(D-3)\Phi Q+(D-2)TS-2VP+2\mathcal{A}\widetilde{\alpha}.
\end{aligned}
\end{equation}

Since there are some problems with holographic interpretion of thermodynamics, we will derive the thermodynamics law for charged Gauss-Bonnet AdS black holes in the next section.

\section{The modified first law of thermodynamics}
Inspired by the AdS/CFT and following the approach in \cite{Cong:2021fnf}, we use the duality relation for the central charge $C$ \cite{Karch:2015rpa} which takes the form as
\begin{equation}
\label{C duality}
\begin{aligned}
C&=k\frac{l^{D-2}}{16\pi G},
\end{aligned}
\end{equation}
where the numerical factor $k$ depends on the details of the particular hologiaphic system. 

We consider $M$ as the function of $M=M(Q,P,J,S,\widetilde{\alpha},G)$ to derive the variation of M
\begin{equation}
\label{variation of M}
\begin{aligned}
\delta M&=\frac{\partial M}{\partial Q}\delta Q+\frac{\partial M}{\partial J}\delta J+\frac{\partial M}{\partial P}\delta P+\frac{\partial M}{\partial S}\delta S+\frac{\partial M}{\partial G}\delta G+\frac{\partial M}{\partial\widetilde{\alpha}}\delta\widetilde{\alpha},
\end{aligned}
\end{equation}
where we define the conjugate variable $\beta$ of $G$ as
\begin{equation}
\label{conjugate variable}
\begin{aligned}
\frac{\partial M}{\partial G}&=-\frac{\beta}{G}.
\end{aligned}
\end{equation}
The Hawking temperature and Bekenstein-Hawking entropy are defined as \cite{Hawking:1974rv,Hawking:1975vcx,Hawking:1976de,Bekenstein:1972tm,Bekenstein:1973ur}
\begin{equation}
\label{temperature and entropy}
\begin{aligned}
T=\frac{\kappa}{2\pi},\quad S=\frac{A}{4G},
\end{aligned}
\end{equation}
where $\kappa$ is the surface gravity and $A$ is the area of black holes. With Eq. (\ref{the pressure and the cosmological constant}) and Eq. (\ref{temperature and entropy}), we can rewrite Eq. (\ref{variation of M}) in terms of $\Lambda$, $A$, $\kappa$ and $\beta$ as
\begin{equation}
\label{rewrite the first law}
\begin{aligned}
\delta M=\Phi\delta Q+\Omega\delta J-\frac{V}{8\pi G}\delta\Lambda+\frac{\kappa}{8\pi G}\delta A-\frac{\beta}{G}\delta G+\mathcal{A}\delta\widetilde{\alpha}.
\end{aligned}
\end{equation}

To calculate $\beta$, we use the modified mass term suggested \cite{Cong:2021fnf}
\begin{equation}
\label{Modified M}
\begin{aligned}
GM&=\mathcal{M}(A,\sqrt{G}Q,GJ,\Lambda,\widetilde{\alpha}),
\end{aligned}
\end{equation}
of which the variation is
\begin{equation}
\label{the variation of modified M}
\begin{aligned}
\delta(GM)&=\frac{\partial\mathcal{M}}{\partial A}\delta A+\frac{\partial\mathcal{M}}{\partial(\sqrt{G}Q)}\delta(\sqrt{G}Q)+\frac{\partial\mathcal{M}}{\partial(GJ)}\delta(GJ)+\frac{\partial\mathcal{M}}{\partial\Lambda}\delta\Lambda+\frac{\partial\mathcal{M}}{\partial\widetilde{\alpha}}\delta\widetilde{\alpha}.
\end{aligned}
\end{equation}
Dividing both sides by $G$, we easily obtain the variation of mass $M$
\begin{equation}
\label{varivaiton of mass M}
\begin{aligned}
\delta M&=\frac{1}{G}(\frac{\partial\mathcal{M}}{\partial A}\delta A+\frac{\partial\mathcal{M}}{\partial\Lambda}\delta\Lambda+\frac{\partial\mathcal{M}}{\partial\widetilde{\alpha}}\delta\widetilde{\alpha})+\frac{\sqrt{G}}{G}\frac{\partial\mathcal{M}}{\partial(\sqrt{G}Q)}\delta Q\\
&+\frac{1}{G}(-M+J\frac{\partial\mathcal{M}}{\partial(GJ)}+\frac{Q}{2\sqrt{G}}\frac{\partial\mathcal{M}}{\partial(\sqrt{G}Q)})\delta G+\frac{\partial\mathcal{M}}{\partial(GJ)}\delta J,
\end{aligned}
\end{equation}
which is different from Eq. (\ref{rewrite the first law}). Comparing the coefficients of the same terms in these two equations, we get 
\begin{equation}
\begin{aligned}
&\frac{\partial\mathcal{M}}{\partial A}=\frac{\kappa}{8\pi},\quad\frac{\partial\mathcal{M}}{\partial\Lambda}=\frac{V}{8\pi},\quad\frac{1}{G}\frac{\partial\mathcal{M}}{\partial\widetilde{\alpha}}=\mathcal{A},\\&\frac{1}{\sqrt{G}}\frac{\partial\mathcal{M}}{\partial(\sqrt{G}Q)}=\Phi,\quad\frac{\partial\mathcal{M}}{\partial(GJ)}=\Omega,
\end{aligned}
\end{equation}
and therefore $\beta$ can be expressed by
\begin{equation}
\label{conjugate variable of G}
\begin{aligned}
\beta&=M-\Omega J+\frac{Q\Phi}{2}.
\end{aligned}
\end{equation}

To introduce the central charge C into the new first law of thermodynamics we are about to deduce, we use Eq. (\ref{the pressure and the cosmological constant}) and  Eq. (\ref{C duality}) to acquire the relation among $\delta G$, $\delta P$ and $\delta C$
\begin{equation}
\label{C G and p}
\begin{aligned}
\frac{\delta G}{G}=-\frac{(D-2)}{D}\frac{\delta P}{P}-\frac{2}{D}\frac{\delta C}{C},
\end{aligned}
\end{equation}
which allows us to recast the first law. In consequence, subsituting $T$, $P$ and $S$ back into Eq. (\ref{rewrite the first law}) when particularly treating $G$ as a variable, we obtain the new first law of thermodynamics for $D$-dimensional Gauss-Bonnet AdS black holes
\begin{equation}
	\label{new first law}
	\begin{aligned}
		\delta M&=\Phi\delta Q+\Omega\delta J+T\delta S+V_{e}\delta P+\mu_{e}\delta C+\mathcal{A}\delta\widetilde{\alpha},
	\end{aligned}
\end{equation}
where we define the effective thermodynamic volume and chemical potential 
\begin{equation}
\label{D effective thermodynamic volume and chemical potential}
\begin{aligned}
V_{e}&=V+\frac{(D-2)\beta}{DP}-\frac{(D-2)(TS+VP)}{D},\\\mu_{e}&=\frac{2\beta}{DC}-\frac{2(TS+VP)}{DC}.
\end{aligned}
\end{equation}

In order to go deeper in the exploration of the formalism above, we derive the expression of $V$ from the smarr relation Eq. (\ref{D smarr relation}) as follows
\begin{equation}
\label{express of DV}
\begin{aligned}
V&=\frac{(D-3)}{2P}\left[\Phi Q+\frac{2\widetilde{\alpha}\mathcal{A}}{(D-3)}-M+\frac{(D-2)}{(D-3)}(\Omega J+TS)\right],
\end{aligned}
\end{equation}
which we substitute into Eq. (\ref{D effective thermodynamic volume and chemical potential}) to simplify the expressions as
\begin{equation}
\label{D effective thermodynamic volume and chemical potential2}
\begin{aligned}
V_{e}&=\frac{2M+4\widetilde{\alpha}\mathcal{A}+(D-4)\Phi Q}{2DP},\quad\mu_{e}&=\frac{2P(V_{e}-V)}{C(D-2)}.
\end{aligned}
\end{equation}

The results above we have obtained are consistent with \cite{Kumar:2022afq}. Next, we will deduce the modified first law for 4-dimensional Gauss-Bonnet AdS black holes and explore the critical behavior and phase transiton structures of $D$-dimensional Gauss-Bonnet AdS black holes.

\section{Critical behaviors and phase transition for $D=4$ and $D=5$}
In this section, we consider the dimension $D=4$ and $D=5$. Due to the added Gauss-Bonnet term, the critical behavior of central charge reveals different properties from \cite{Cong:2021fnf}. Then we further study the phase transition structures based on the varivation of central charge.

\subsection{D=4}
Recently, $4D$ charged Gauss-Bonnet black holes attract a lot of attention \cite{Glavan:2019inb,Fernandes:2020rpa,Wei:2020poh}. Considering the limit $D\rightarrow 4$, we obtain the black holes solution \cite{Fernandes:2020rpa,Wei:2020poh}
\begin{equation}
\begin{aligned}
ds^{2}&=-f(r)dt^{2}+\frac{1}{f(r)}dr^{2}+r^{2}(d\theta^{2}+sin^{2}\theta d\phi^{2}),
\end{aligned}
\end{equation}
where the metric function is 
\begin{equation}
\label{the 4 metric function}
\begin{aligned}
f(r)&=1+\frac{r^{2}}{2\alpha}(1-\sqrt{1+4\alpha(\frac{2MG}{r^{3}}-\frac{Q^{2}G}{r^{4}}-\frac{1}{l^{2}})}.
\end{aligned}
\end{equation}
The first law of thermodynamics for 4-dimensional charged Gauss-Bonnet AdS black holes is 
\begin{equation}
\label{4 dimensional first law}
\begin{aligned}
\delta M&=T\delta S+V\delta P+\Phi\delta Q+\mathcal{A}_{4}\delta\alpha,
\end{aligned}
\end{equation}
where $\mathcal{A}_4$ is the conjugate variable of $\alpha$ in $D=4$, and it is easy to confirm the smarr relation
\begin{equation}
\begin{aligned}
M&=\Phi Q+2TS-2VP+2\mathcal{A}_{4}\alpha.
\end{aligned}
\end{equation}

Following the same step in section II, we obtain the modified first law of 4-dimensional charged Gauss-Bonnet AdS black holes
\begin{equation}
\label{first law 4Dgauss}
\begin{aligned}
\delta M&=\Phi\delta Q+\Omega\delta J+T\delta S+V_{4e}\delta P+\mu_{4e}\delta C+\mathcal{A}_{4}\delta\alpha,
\end{aligned}
\end{equation}
where
\begin{equation}
\label{V and mu 4D gauss}
\begin{aligned}
V_{4e}&=V+\frac{\beta}{2P}-\frac{(TS+VP)}{2}=\frac{M+2A_{4}\alpha}{4P},\\\mu_{4e}&=\frac{\beta}{2C}-\frac{(TS+VP)}{2C}=\frac{P(V_{c}-V)}{C},
\end{aligned}
\end{equation}
are black holes effective thermodynamic volume and chemical potential.

The temperature of 4-dimensional charged Gauss-Bonnet AdS black holes is 
\begin{equation}
\begin{aligned}
T&=\frac{3r_{+}^{4}+l^{2}r_{+}^{2}-Gl^{2}Q^{2}-\alpha l^{2}}{4\pi l^{2}(2\alpha r_{+}+r_{+}^{3})},
\end{aligned}
\end{equation}
and the critical value can be obtained by \cite{Cong:2021fnf}
\begin{equation}
\label{critical value}
\begin{aligned}
\frac{\partial T}{\partial r_{+}}=0,\quad\frac{\partial^{2}T}{\partial r_{+}^{2}}=0,
\end{aligned}
\end{equation}
which allows us to obtain the critical value of $r$, $l$ and $T$ as follows
\begin{equation}
\label{critical value D4}
\begin{aligned}
r_{c}&=\sqrt{X_{2}+6\alpha},\\
l_{c}&=\frac{6\sqrt{X_{3}}}{\sqrt{7\alpha\left(X_{1}+7GQ^{2}\right)+3GQ^{2}X_{2}+52\alpha^{2}}},\\
T_{c}&=\frac{2X_{4}}{\pi\sqrt{X_{2}+6\alpha}\left(X_{2}+8\alpha\right)X_{3}},
\end{aligned}
\end{equation}
where
\begin{equation}
\begin{aligned}
X_{1}&=\sqrt{48\alpha^{2}+48\alpha GQ^{2}+9G^{2}Q^{4}},\\X_{2}&=X_{1}+3GQ^{2},\\X_{3}&=132\alpha^{3}+3G^{2}Q^{4}X_{2}+8\alpha GQ^{2}\left(2X_{1}+9GQ^{2}\right)+\alpha^{2}\left(19X_{1}+177GQ^{2}\right),\\X_{4}&=2\left(284\alpha^{4}+\alpha G^{2}Q^{4}\left(23X_{1}+93GQ^{2}\right)+11\alpha^{2}GQ^{2}\left(5X_{1}+31GQ^{2}\right)\right)\\&+3G^{3}Q^{6}X_{2}+\alpha^{3}\left(41X_{1}+523GQ^{2}\right).
\end{aligned}
\end{equation}
We also use the definition of the central charge Eq. (\ref{C duality}) to obtain the critical value 
\begin{equation}
\label{critical value C 4D}
\begin{aligned}
C_{c}=k\frac{l_{c}^{2}}{16\pi G}=\frac{9k\left(132\alpha^{3}+3G^{2}Q^{4}X_{2}+8\alpha GQ^{2}\left(2X_{1}+9GQ^{2}\right)+\alpha^{2}\left(19X_{1}+177GQ^{2}\right)\right)}{\left(4\pi G\left(7\alpha\left(X_{1}+7GQ^{2}\right)+3GQ^{2}X_{2}+52\alpha^{2}\right)\right.}.
\end{aligned}
\end{equation}
When the Gauss-Bonnet coupling constant $\alpha\rightarrow0$, critical values of $l$, $r$, $T$ and central charge $C$ will reduce as follows
\begin{equation}
\label{critical value without alpha}
\begin{aligned}
r_{c}&\rightarrow\sqrt{6G}Q,\quad l_{c}\rightarrow6\sqrt{G}Q,\quad T_{c}\rightarrow\frac{\sqrt{6}}{18\sqrt{G}\pi Q},\quad C_{c}\rightarrow\frac{9kQ^{2}}{4\pi},
\end{aligned}
\end{equation}
which are consistent with that of asymptotically AdS black holes \cite{Cong:2021fnf}. From Eq. (\ref{critical value without alpha}), we find that the critical value of the central charge is universal, i.e. it is only determined by $Q$ regardless of the value of $P$. But when considering the Gauss-Bonnet gravity, Eq. (\ref{critical value C 4D}) reveals that the critical value is also determined by $\alpha$. Next, we will study the phase transition structures based on different values of central charge $C$.

The mass and entropy of 4-dimensional charged Gauss-Bonnet AdS black holes are \cite{Fernandes:2020rpa,Wei:2020poh}
\begin{equation}
\label{mass and entropy}
\begin{aligned}
M=\frac{\alpha l^{2}+r_{+}^{4}+l^{2}r_{+}^{2}+Gl^{2}Q^{2}}{2Gl^{2}r_{+}},\quad S=\frac{1}{G}(\pi r_{+}^{2}+4\pi\alpha ln(\frac{r_{+}}{\sqrt{\alpha}})),
\end{aligned}
\end{equation}
and the free energy is
\begin{equation}
\label{free energy D4}
\begin{aligned}
F&=M-TS\\
 &=\frac{\left(4\alpha ln\left(\frac{r_{+}}{\sqrt{\alpha}}\right)+r_{+}^{2}\right)\left(\alpha+G\left(Q^{2}-8\pi Pr_{+}^{4}\right)-r_{+}^{2}\right)}{4r_{+}G\left(2\alpha+r_{+}^{2}\right)}+\frac{\left(\alpha+r_{+}^{2}\right)}{2r_{+}G}+\frac{4\pi Pr_{+}^{3}}{3}+\frac{Q^{2}}{2r_{+}}.
\end{aligned}
\end{equation}

\begin{figure}[H]
	\centering
	\includegraphics[scale=0.42]{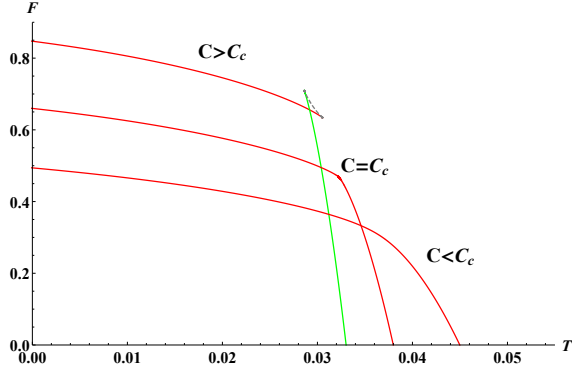}
	\caption{The Gibbs free energy $F$ with  temperature $T$ for $4D$ charged Gauss-Bonnet AdS black holes. We fix $Q=0.5$, $\alpha=0.5$, $k=16\pi$, $P=0.0014$. There exists one critical point $C_c=42.9628$. The red/green soild lines indicate the thermodyanmically stable branches. The thermodyanmically unstable branch is represented by the gray dashed line. We take the values of central charge $C$ as 80, $C_c$ and 40 from top to bottom.}
	\label{free energy fix the central charge C 4D}
\end{figure}

In order to show the behavior of the free energy, we describe it in Fig. (\ref{free energy fix the central charge C 4D}) for different values of the central charge. 

For this case, there is only one critical point of central charge $C_c=42.9628$. The Fig. (\ref{free energy fix the central charge C 4D}) shows that for fixed value of pressure $P$. When $C<C_c$, there is only a smooth curve, which represents a thermodynamically stable branch indiacting no phase transition structures. For the case of $C=C_c$, the system is in critical state, from which phase transition behavior occurs in the region above. When the central charge continues to increase to $C_c$, another thermodynamically stable branch occurs. There are two stable branches and one unstable branch represented by the gray dashed line. There is a swallowtail structure to express the first order phase transiiton structures of small/large black holes.

\subsection{D=5}
From Eq. (\ref{temperature1}), the Hawking temperature of charged AdS Gauss-Bonnet black holes in $D=5$ is
\begin{equation}
\label{Hawking T D5}
\begin{aligned}
T&=\frac{\frac{4r_{+}^{4}}{l^{2}}+2r_{+}^{2}-\frac{2Q^{2}G}{3r_{+}^{2}}}{4\pi r_{+}(r_{+}^{2}+2\widetilde{\alpha})}.
\end{aligned}
\end{equation}
Substituting the equation above into Eq. (\ref{critical value}) to calculate the critical value of $l$ and using Eq. (\ref{C duality}), we can obtain the critical value of central charge $C_c$. We find that it is too hard to obtain the analytic solution of $C_c$. The critical value of central charge $C$ can be obtained by expanding $\alpha$ perturbatively at the first order.  From previous analysis, the critical central only depends on the charge $Q$ and is called the universal property. The universal nature of the central charge is broken due to the existence of $\alpha$ as shown in \cite{Kumar:2022afq}. Here, we choose to numerically analyse critical central charge and phase transition structures. 

From Eq. (\ref{entropy1}), the entropy of 5-dimensional charged Gauss-Bonnet AdS black holes is
\begin{equation}
\begin{aligned}
S&=\frac{r_{+}^{3}}{4G}(1+\frac{6\widetilde{\alpha}}{r_{+}^{2}}),
\end{aligned}
\end{equation}
therefore the free energy of black holes is
\begin{equation}
	\label{free energy D5}
	\begin{aligned}
		F=\frac{18\widetilde{\alpha}^{2}l^{2}r_{+}^{2}+18\widetilde{\alpha}Gl^{2}Q^{2}-9\widetilde{\alpha}l^{2}r_{+}^{4}-54\widetilde{\alpha}r_{+}^{6}+5Gl^{2}Q^{2}r_{+}^{2}+3l^{2}r_{+}^{6}-3r_{+}^{8}}{96\pi\widetilde{\alpha}Gl^{2}r_{+}^{2}+48\pi Gl^{2}r_{+}^{4}}.
	\end{aligned}
\end{equation}

\begin{figure}[H]
	\centering
	\includegraphics[scale=0.42]{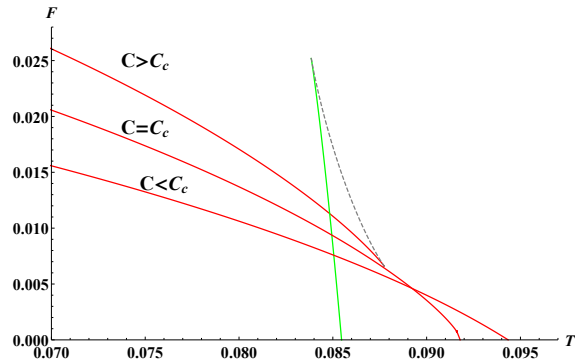}
	\caption{The Gibbs free energy $F$ with temperature $T$ for $5D$ charged Gauss-Bonnet AdS black holes. We use $Q=0.18$, $\widetilde{\alpha}=0.5$, $k=16\pi$, $P=0.015$ and the critical point for this case is $C_c=88.3795$. Thermodynamically stable branches are represented by red/green solid lines while the gray dashed line indicates the unstable branch. The central charge $C$ is fixed as 170, $C_c$ and 40 respectively.}
	\label{f-T 5D}
\end{figure}

We can plot the diagram of Gibbs free energy of black holes with temperature in Fig. (\ref{f-T 5D}).

For 5-dimensional charged Gauss-Bonnet AdS black holes, we also plot the $F-T$ diagram in Fig. (\ref{f-T 5D}). We can obtain the critical central charge $C_c=88.3795$ with above parameters. The diagram shows that when $C< C_c$, the red solid curve is smooth which indicates that is no phase transition behavior, and the case $C=C_C$ still shows the same phase transition as $C<C_c$. A swallowtail appears to show the first order phase transition structures of small/large black holes for $C_c<C$. The phase transition structures in this case are same as 4-dimensional black holes and consistent with \cite{Kumar:2022afq}.

\section{The triple points and phase transition structures for $D=6$}
For 6-dimensional charged Gauss-Bonnet AdS black holes, from Eq. (\ref{temperature1}) and (\ref{entropy1}), the temperature and entropy are
\begin{equation}
\label{temperature 6D}
\begin{aligned}
T=\frac{\frac{5r_{+}^{4}}{l^{2}}+3r_{+}^{2}+\widetilde{\alpha}-\frac{Q^{2}G}{2r_{+}^{4}}}{4\pi r_{+}(r_{+}^{2}+2\widetilde{\alpha})}, 
\end{aligned}
\end{equation}

\begin{equation}
\begin{aligned}
S&=\frac{r_{+}^{4}}{4G}(1+\frac{4\widetilde{\alpha}}{r_{+}^{2}}),
\end{aligned}
\end{equation}
and the free energy of 6-dimensional charged Gauss-Bonnet AdS black holes is
\begin{equation}
\label{free energy D6}
\begin{aligned}
F&=\frac{24\widetilde{\alpha}^{2}l^{2}r_{+}^{4}+20\widetilde{\alpha}Gl^{2}Q^{2}-6\widetilde{\alpha}l^{2}r_{+}^{6}-72\widetilde{\alpha}r_{+}^{8}+7Gl^{2}Q^{2}r_{+}^{2}+6l^{2}r_{+}^{8}-6r_{+}^{10}}{192\pi\widetilde{\alpha}Gl^{2}r_{+}^{3}+96\pi Gl^{2}r_{+}^{5}}.
\end{aligned}
\end{equation}

There are more complicated phase transiton structures and the triple point of charged Gauss-Bonnet AdS black holes based on the varivation of $P$ for some parameters in $D=6$ \cite{Wei:2014hba}, and thus we will study whether there are more complex phase transition behaviors based on the varivation of $C$.

First, we discuss about some properties of critical central charge. The critical point can be obtained with Eq. (\ref{critical value}).  We plot the diagram of central charge $C$ with charge $Q$.
\begin{figure}[H]
    \centering
    \includegraphics[scale=0.42]{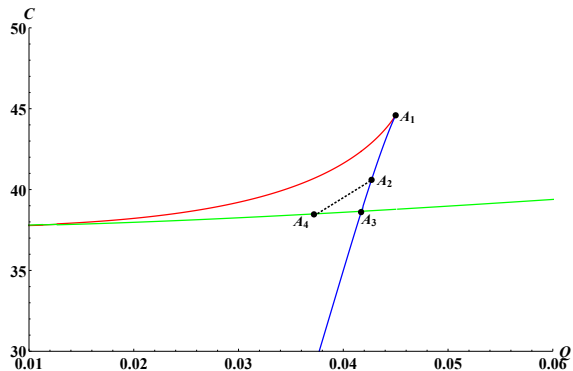}
	\caption{$C-Q$ diagram of $6D$ charged Gauss-Bonnet AdS black holes. We fix $k=16\pi$, $P=0.015$, $\widetilde{\alpha}=0.5$. The blue, red and green solid lines indicate different values of critical central charge. The coordinates of the points are $A_1=(0.045,44.592)$, $A_2=(0.042,38.6064)$, $A_3=(0.0427,40.6016)$ and $A_4=(0.0372,38.4664)$. The line linking $A_2$ and $A_4$ represents different triple points for different values of charge $Q$.} 
	\label{C-Q 6D}
\end{figure}
For different values of $Q$, there are different numbers of critical points. From Fig. (\ref{C-Q 6D}), we can see that there are three critical points when $Q<Q_3=0.045$, while the triple point occurs when $Q_1=0.0372<Q<Q_2=0.0427$. All these imply the existence of complex phase structures, which will be discussed separately. In the following discussion, we take parameters $\widetilde{\alpha}=0.5$, $P=0.015$ and $k=16\pi$.

\paragraph{$Q<Q_1$}
~{}

\begin{figure}[H]
	\subfigure[$C<C_\alpha$]{
	\begin{minipage}[t]{0.5\textwidth}
			\centering
			\includegraphics[scale=0.37]{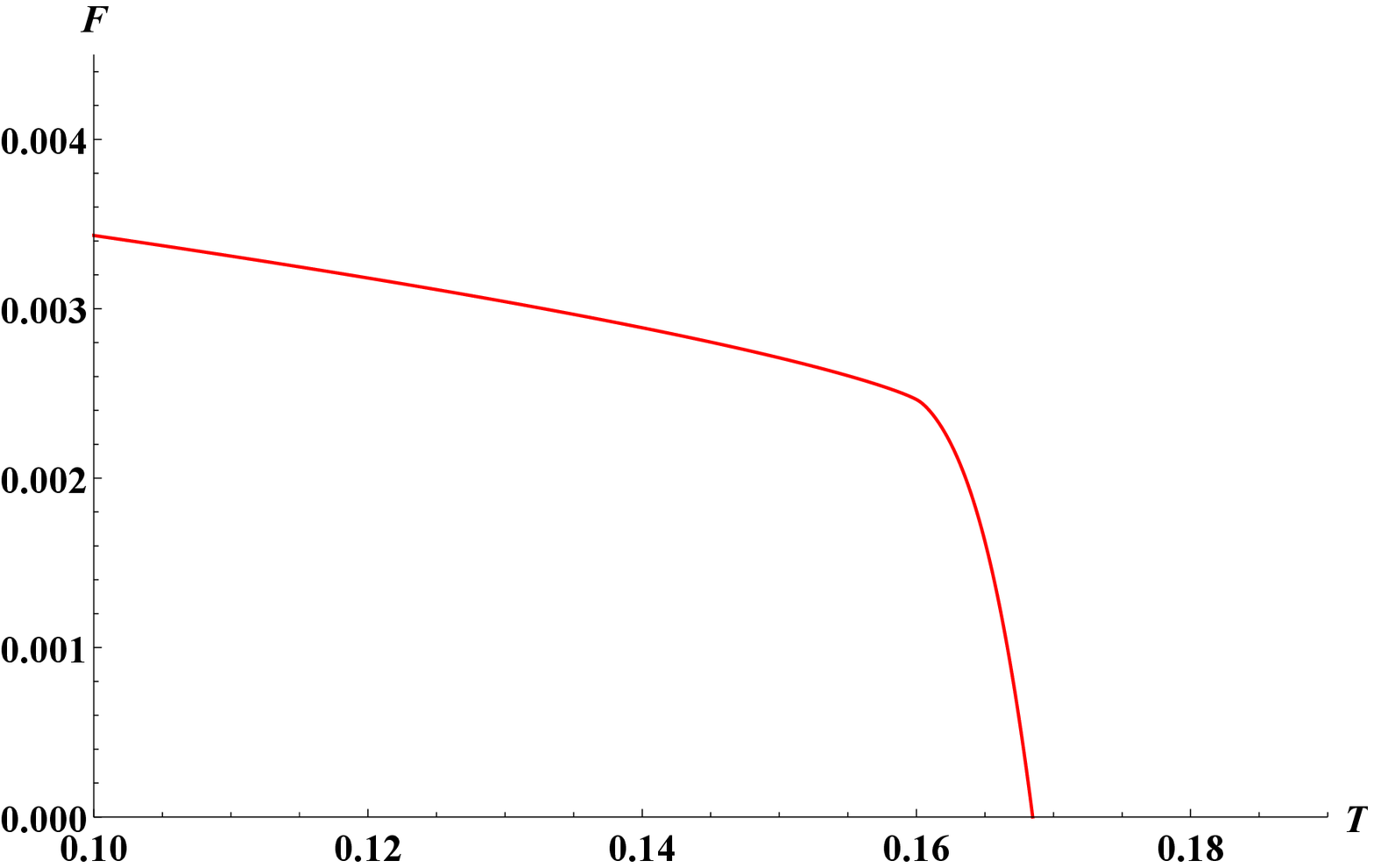}
			\label{C=20}
	\end{minipage}}
	\subfigure[$C_\alpha<C<C_\beta$]{
	\begin{minipage}[t]{0.5\textwidth}
		\centering
		\includegraphics[scale=0.37]{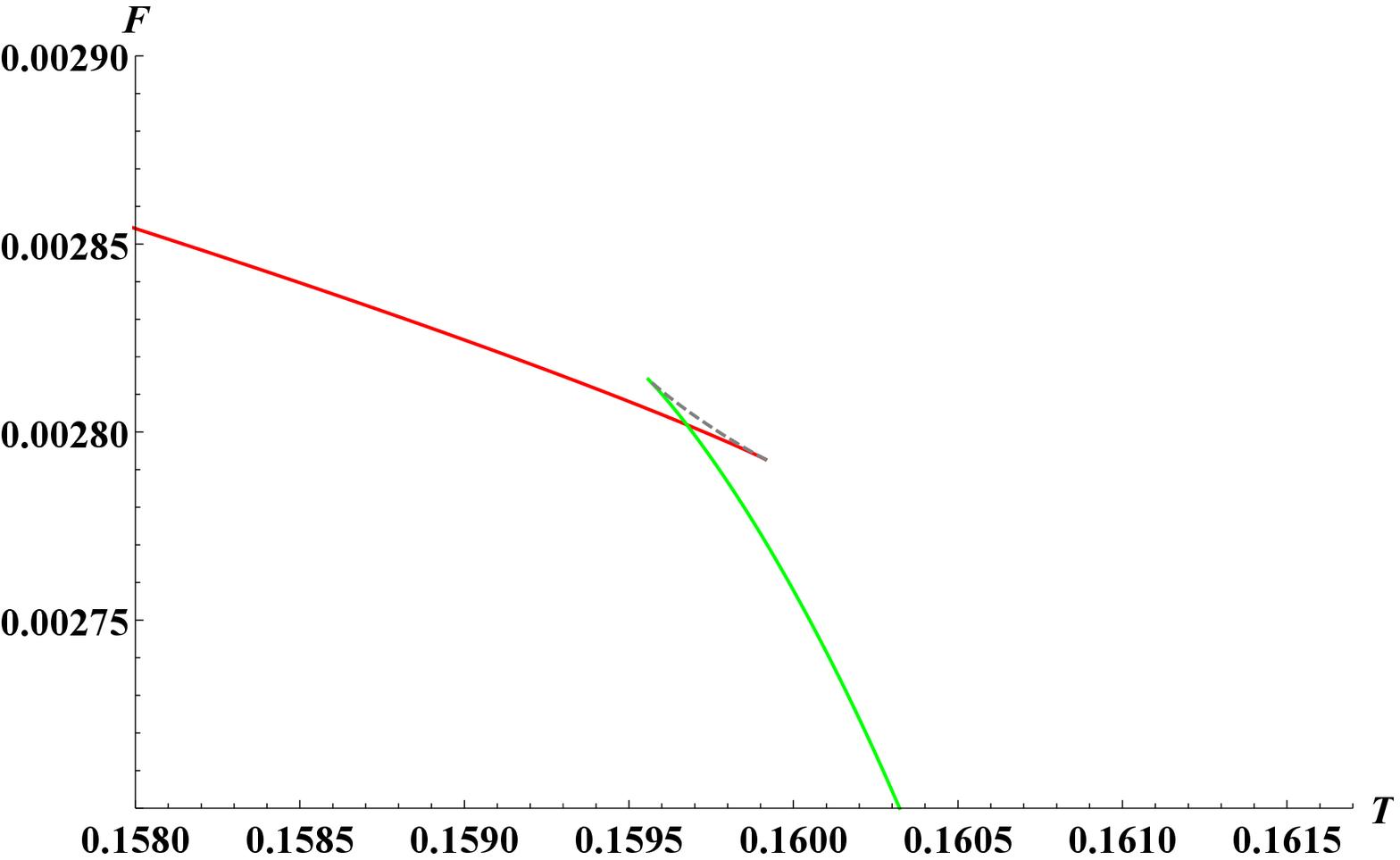}
		\label{C=30}
\end{minipage}}
	\subfigure[$C_\beta<C<C_\gamma$]{
		\begin{minipage}[t]{0.5\textwidth}
		\centering
		\includegraphics[scale=0.37]{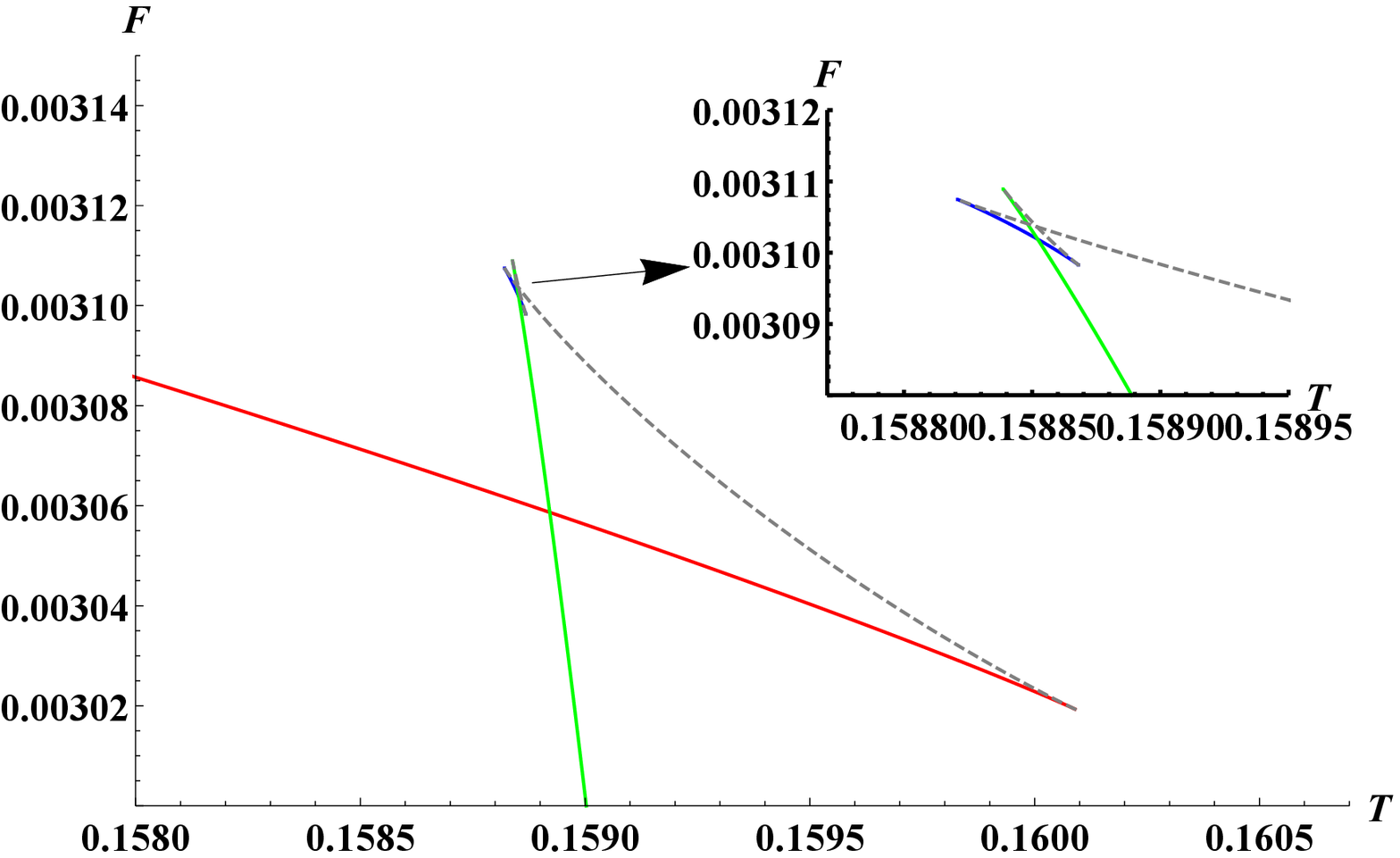}
		\label{C=39}
	\end{minipage}}
	\subfigure[$C>C_\gamma$]{
		\begin{minipage}[t]{0.5\textwidth}
	     \centering
         \includegraphics[scale=0.37]{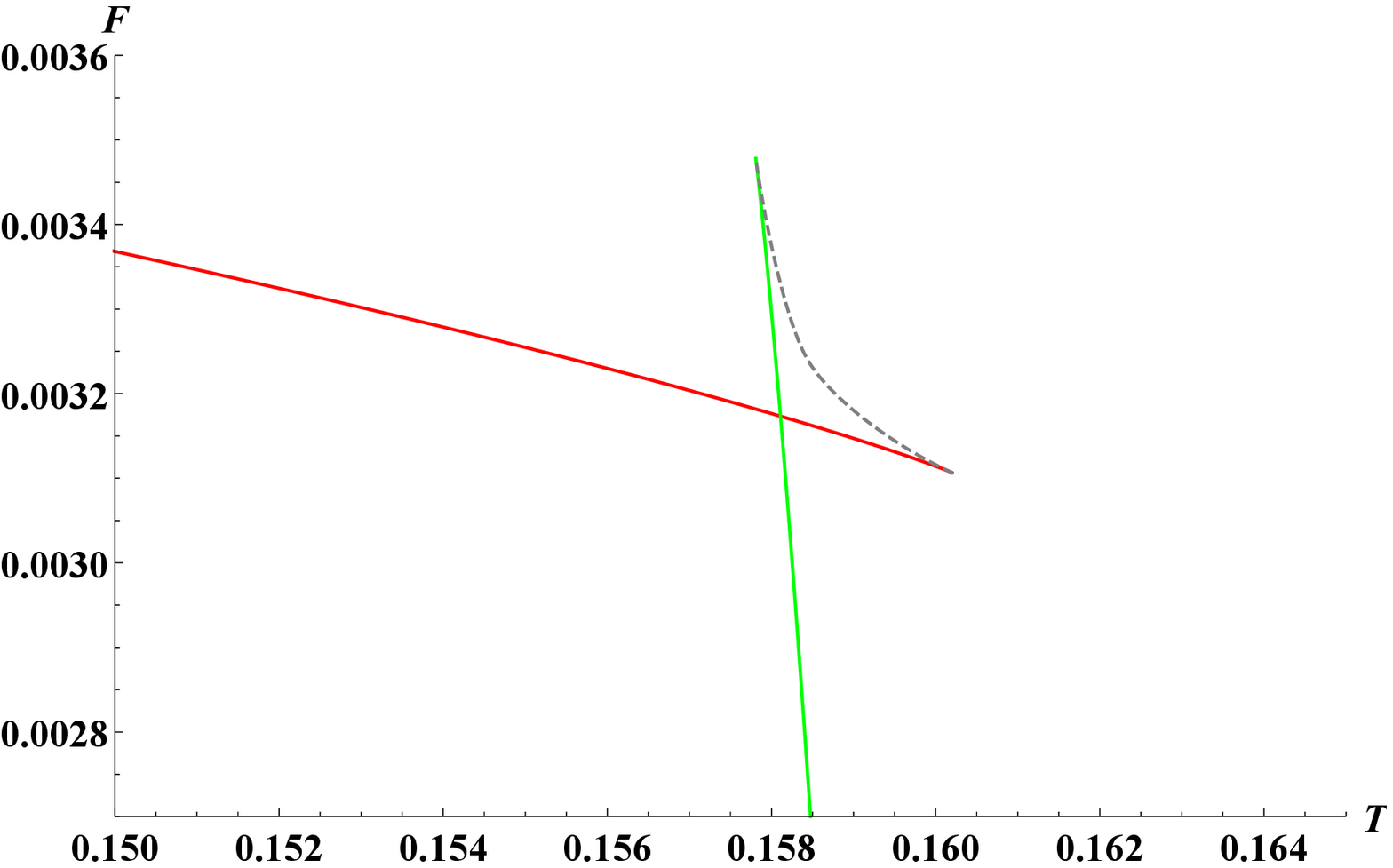}
		\label{C=43}
	\end{minipage}}
    \caption{The Gibbs free energy $F$ with temperature $T$ for $Q=0.035$ for $6D$ charged Gauss-Bonnet AdS black holes. There are three critical points $(C_\alpha,C_\beta,C_\gamma)$, and therefore we take the value of central charge $C$ as 20, 30, 39 and 43 respectively. The gray dashed lines describes thermodynamically unstable branches, while the red/green/blue solid lines represent stable ones.}
    \label{Q=0.035}
\end{figure}

In this case, without loss of generality, we take $Q=0.035$. There are three critical points of central charge $C_{\alpha}=24.3138$, $C_{\beta}=38.4128$ and $C_{\gamma}=40.13$. Based on these critical points, we take four different values of $C$ to plot diagrams of Gibbs free energy $F$ with temperature $T$ as shown in Fig. (\ref{Q=0.035})

Fig. (\ref{C=20}) shows that when $C<C_{\alpha}$, the $F-T$ diagram is only a smooth curve which indicates no phase transition structure. When $C_{\alpha}<C$, there are two thermodynamically stable branches representing small black holes and large black holes respectively and an unstable intermediate black holes branch. There is only a swallowtail behavior that represents the first order phase transition structures of small/large black holes in Fig. (\ref{C=30}). When the cenrtal charge continues to increase until it exceeds $C_{\beta}$, a new swallowtail appears. It indicates the appearance of a new stable intermediate black holes branch described by the blue solid line in Fig. (\ref{C=39}). The diagram still exhibits the small/large black holes phase transition for  $C_{\beta}<C<C_{\gamma}$ due to large value of $F$, which does not affect the phase transition. Although there is no new phase transition structure in this case, we can observe the richer structure of $F-T$ diagram from Fig. (\ref{C=39}). When $C>C_{\gamma}$, the new swallowtail disappearing, and the situation in Fig .(\ref{C=43}) still shows the same small/large black holes phase transition structures. 
\paragraph{$Q_1<Q<Q_2$}
~{}

For this case, there exist three critical points. In general, we take $Q=0.04$, and therefore three critical central charge are $(C_{\alpha},C_{\beta},C_{\gamma})=(34.9884,38.5938,41.6213)$. 

\begin{figure}[H]
		\subfigure[$C<C_\alpha$ and $C_\alpha<C<C_\beta$]{
		\begin{minipage}[t]{0.5\textwidth}
		\centering
		\includegraphics[scale=0.37]{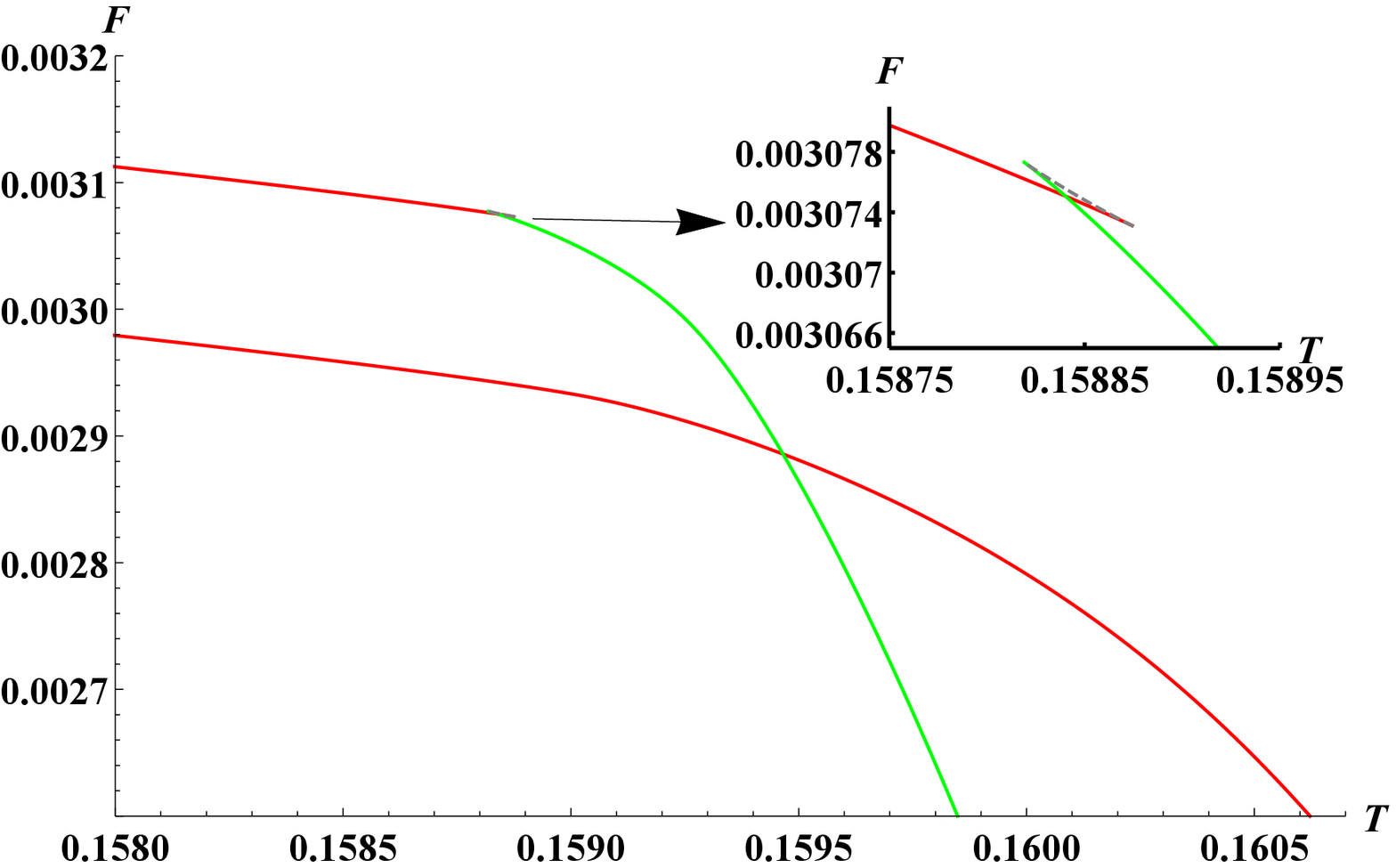}
		\label{C=32 37}
	    \end{minipage}}
	\subfigure[$C_\beta<C<C_t$]{
		\begin{minipage}[t]{0.5\textwidth}
		\centering
		\includegraphics[scale=0.37]{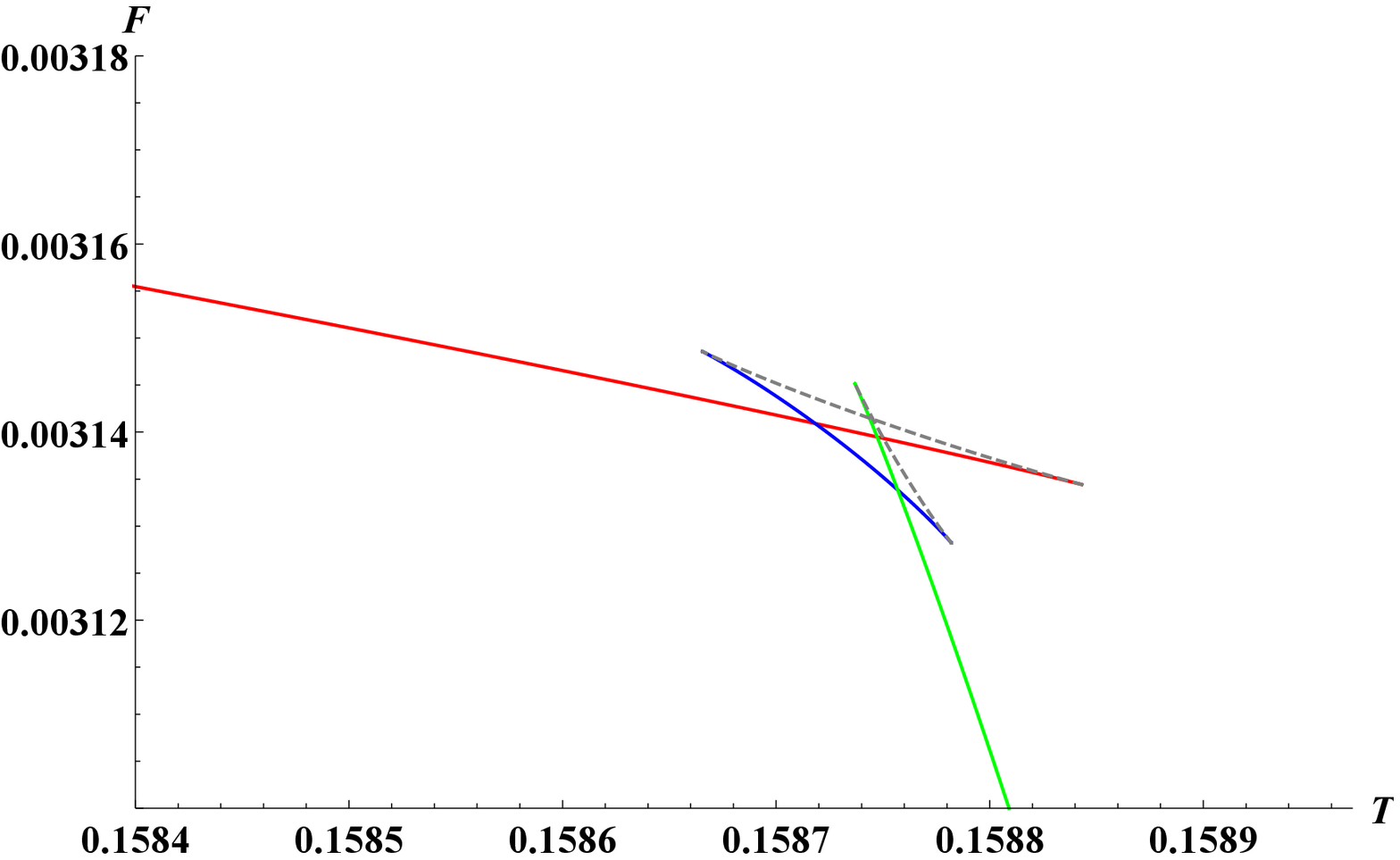}
		\label{C=39.4}
	\end{minipage}}
        \subfigure[$C_t<C<C_\gamma$]{
        \begin{minipage}[t]{0.5\textwidth}
		\centering
		\includegraphics[scale=0.37]{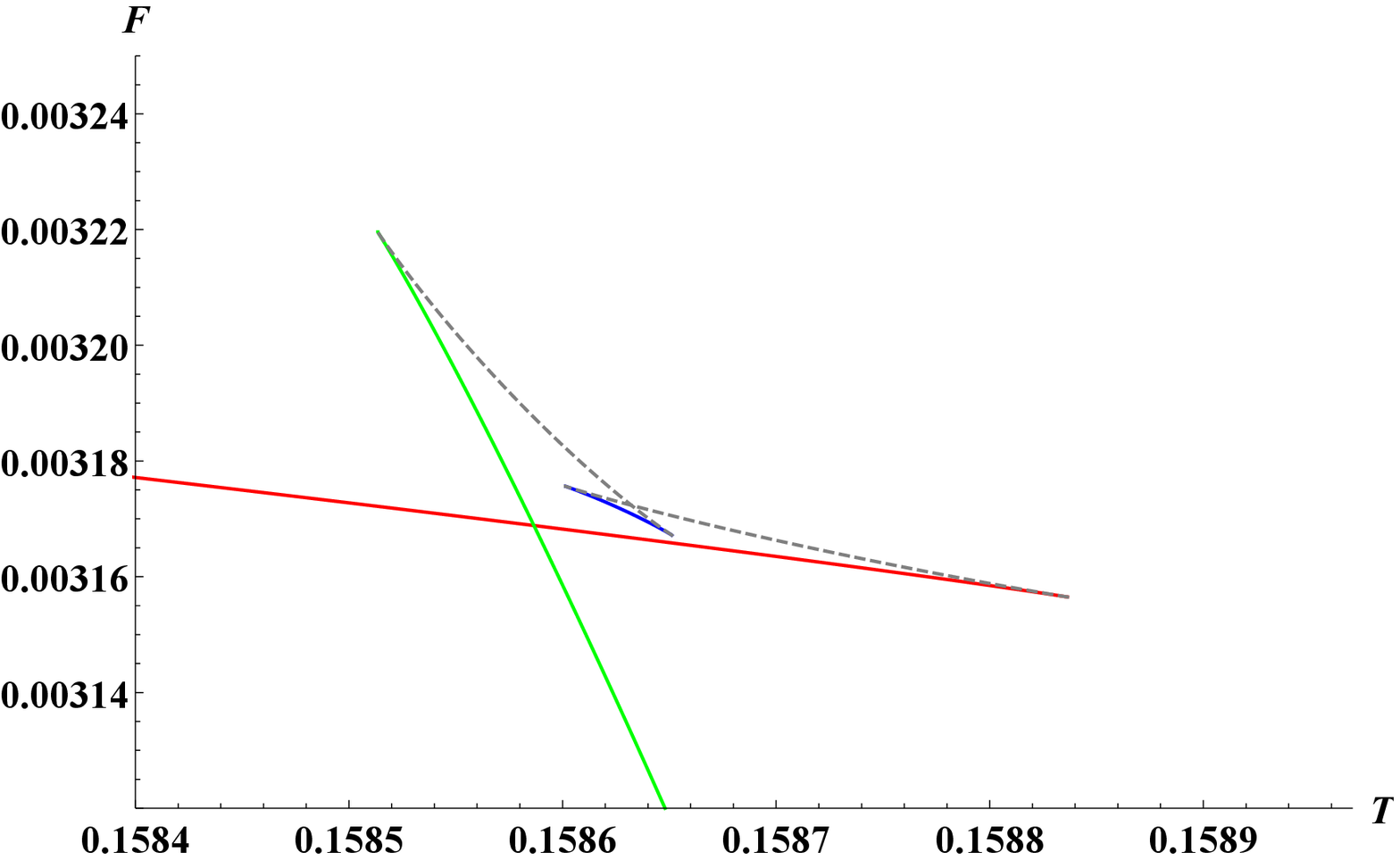}
		\label{C=40.3}	
	    \end{minipage}}
		\subfigure[$C>C_\gamma$]{
		\begin{minipage}[t]{0.5\textwidth}
		\centering
		\includegraphics[scale=0.37]{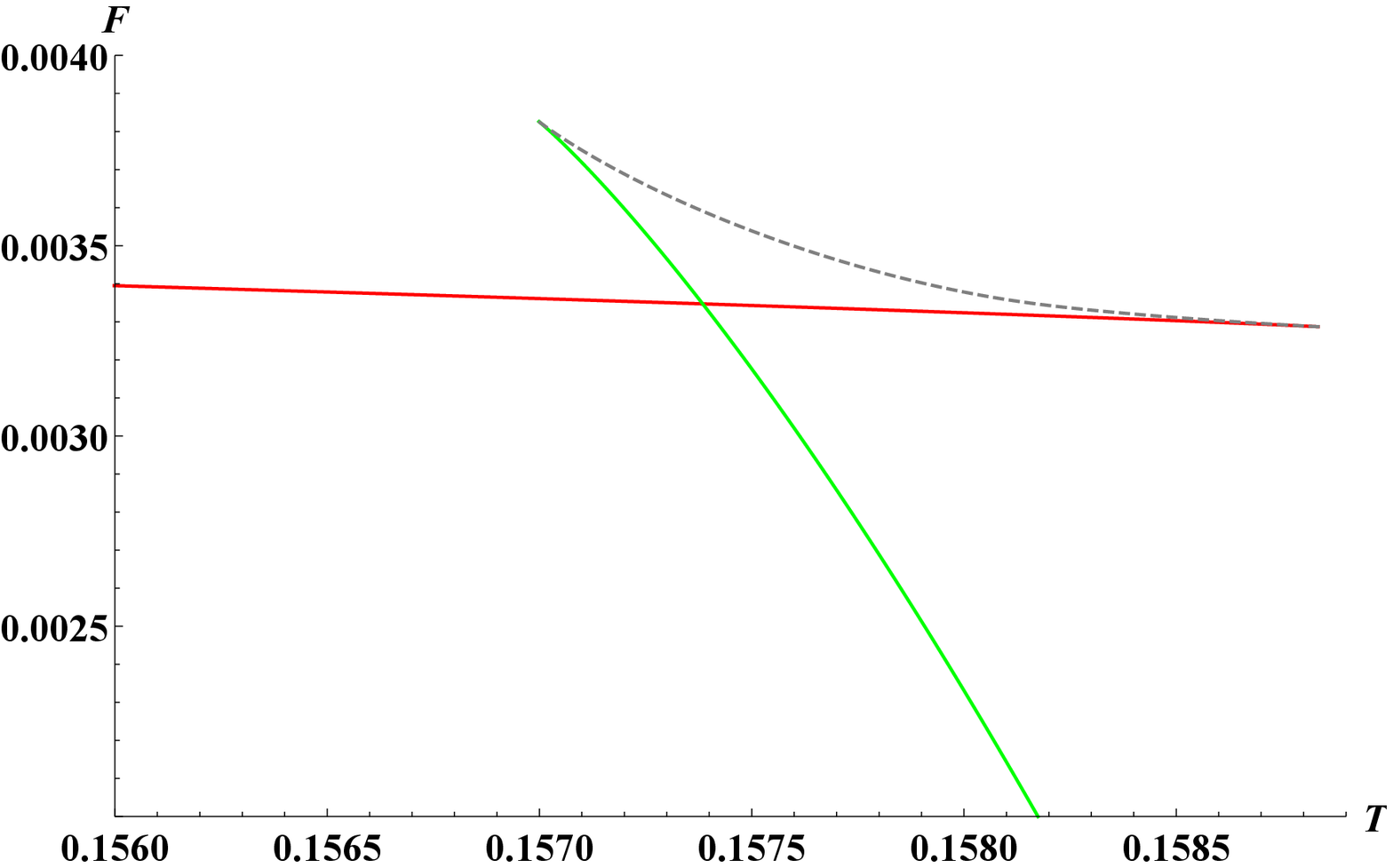}
		\label{C=46}
		\end{minipage}}
   \caption{The Gibbs free energy $F$ with temperature $T$ for $Q=0.04$. The critical central charge are $(C_{\alpha},C_{\beta},C_{\gamma})$, and there exists a triple point $C_t$, and t. Therefore, we take $C$ as 32, 37, 39.4, 40.3 and 46 respectively to analyse the different cases of phase transition structures. The red/green/blue solid lines represent the thermodynamically stable branches, and the gray dashed lines represent the unstable branches. }
   \label{Q=0.04 1}
\end{figure}

In Fig. (\ref{C=32 37}), there is a smooth curve indicating no phase transition structure when $C<C_{\alpha}$ and a swallowtail behavior exhibiting the first order phase transition structure of small/large black holes when $C_{\alpha}<C<C_{\beta}$, as same as our previous analysis in Fig. ({\ref{C=20}}) and Fig. (\ref{C=30}). As for $C_{\beta}<C<C_{\gamma}$, a new thermodynamically stable branch described by the blue solid line appears, and phase transition behavior has become more complicated than ever. For $C>C_\gamma$ as shown in Fig. (\ref{C=46}), the new branch disappears. There are only two thermodynamically stable branches and one swallowtail behavior demonstrating the same first order phase transition structures of small/large black holes as in Fig. (\ref{C=43}).

\begin{figure}[H]
	\centering
	\includegraphics[scale=0.42]{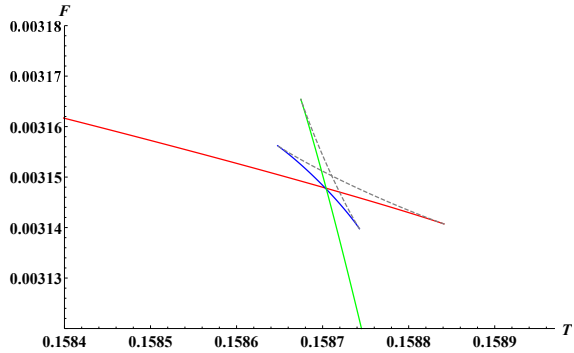}
	\caption{The Gibbs free energy $F$ with temperature $T$ for $C=C_t=39.655$. For this case, a new thermodynamically stable branch described by the blue line appears, and therefore there exist three thermodynamically stable branches. The triple point occurs, where the red/green/blue solid lines intersect, and three thermodynamically stable branches coexist. }
	\label{C=39.655}
\end{figure} 

For $C_\beta<C<C_\gamma$, there exsits the triple point $C_t=39.655$. When the value of central charge exceeds $C_{\beta}$, the blue line appears, indicating a thermodynamically stable intermediate black holes branch. For the case $C_\beta<C<C_{t}$, there are two swallowtails to exhibit small/intermediate/large black holes phase transition as shown in Fig. (\ref{C=39.4}). In this case, the phase transition structures are second-order. With $C$ approaching $C_t$, three branches gradually intersect at one point, which is the triple point $C=C_t$ in Fig. (\ref{C=39.655}), where small, intermediate and large black holes can coexist. For $C_{t}<C<C_\gamma$, although the stable intermediate black holes branch and swallowtail still exsit in Fig. (\ref{C=40.3}), there is only the first order phase transition of small/large black holes.  
 
\paragraph{$Q_2<Q<Q_3$}
~{}

\begin{figure}[H]
		\subfigure[$C<C_\alpha$]{
		\begin{minipage}[t]{0.5\textwidth}
			\centering
			\includegraphics[scale=0.37]{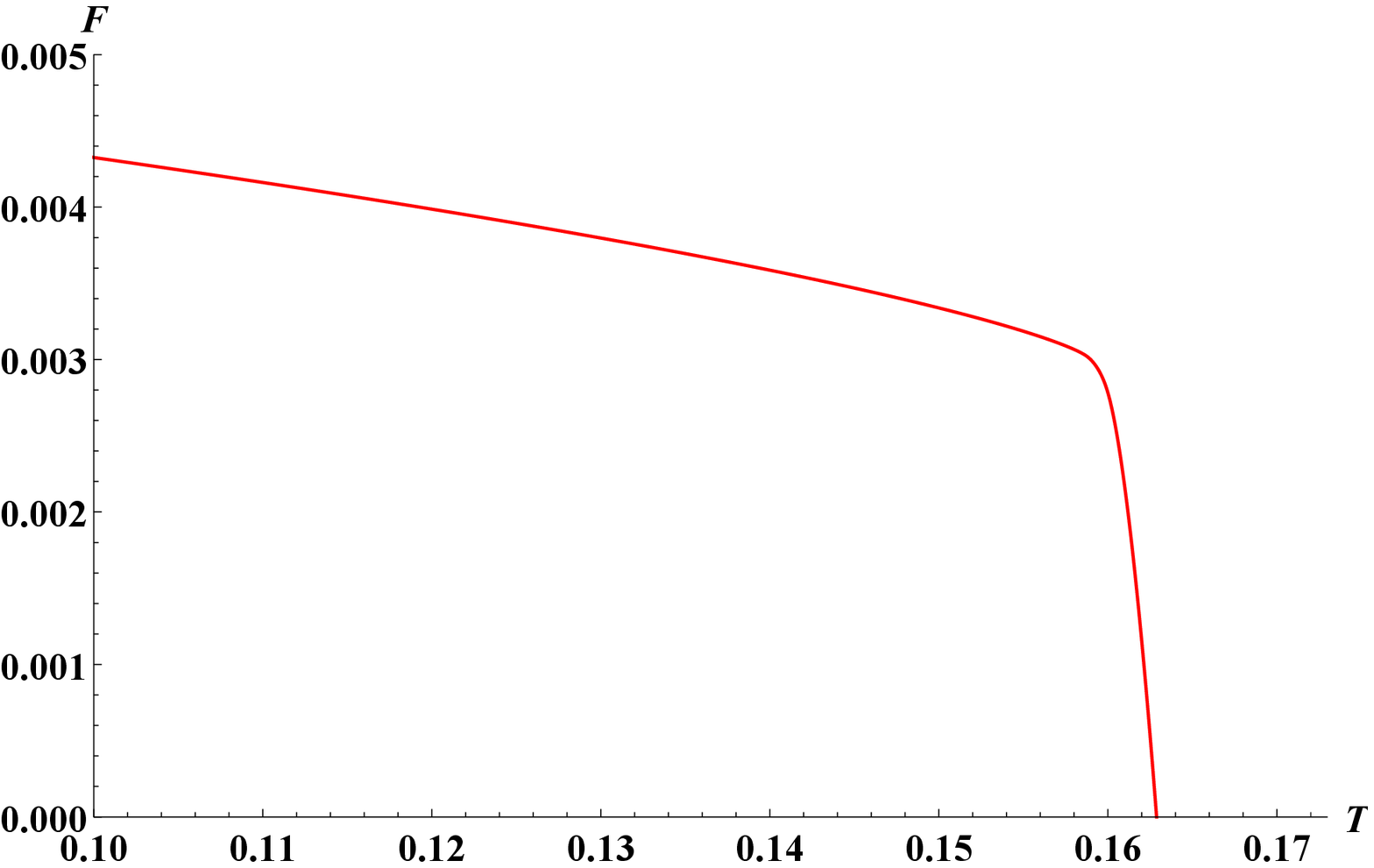}
			\label{C=34}
	\end{minipage}}
       \subfigure[$C_\alpha<C<C_\beta$]{
       \begin{minipage}[t]{0.5\textwidth}
        	\centering
        	\includegraphics[scale=0.37]{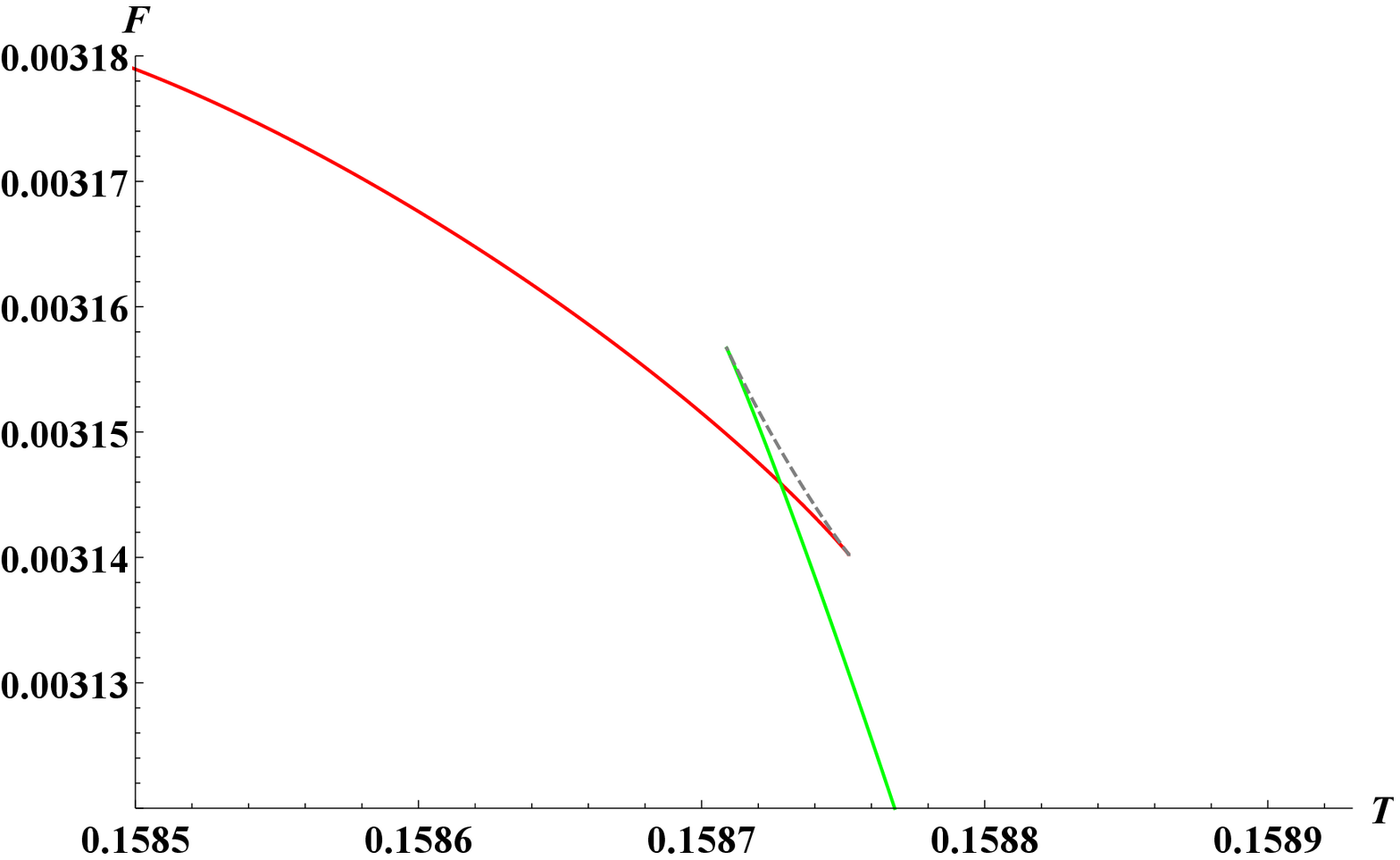}
        	\label{C=39.5}
        \end{minipage}}
	    \subfigure[$C_\beta<C<C_\gamma$]{
		\begin{minipage}[t]{0.5\textwidth}
		\centering
		\includegraphics[scale=0.37]{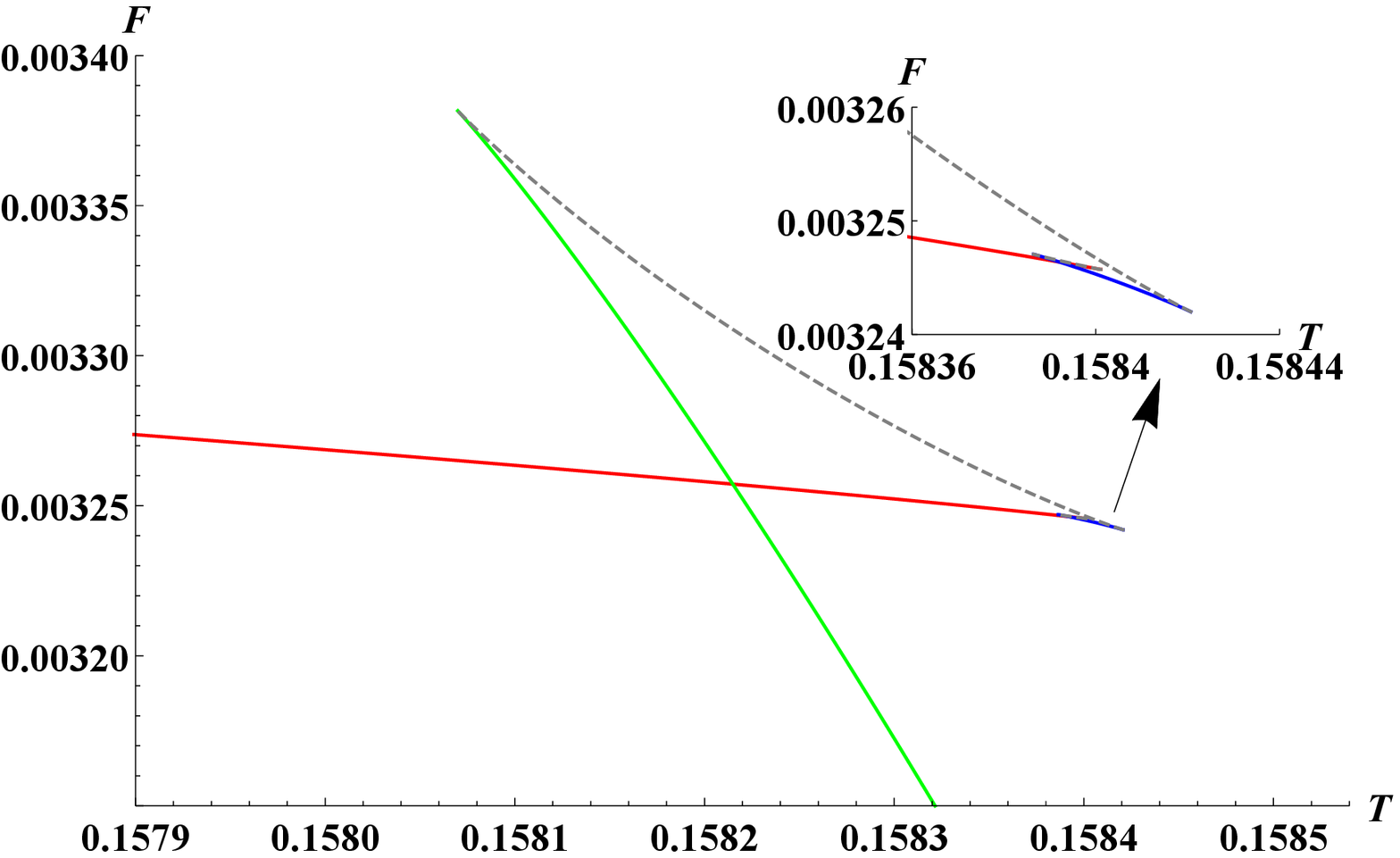}
		\label{C=42}
	\end{minipage}}
        \subfigure[$C>C_\gamma$]{\begin{minipage}[t]{0.5\textwidth}
		\centering
        \includegraphics[scale=0.37]{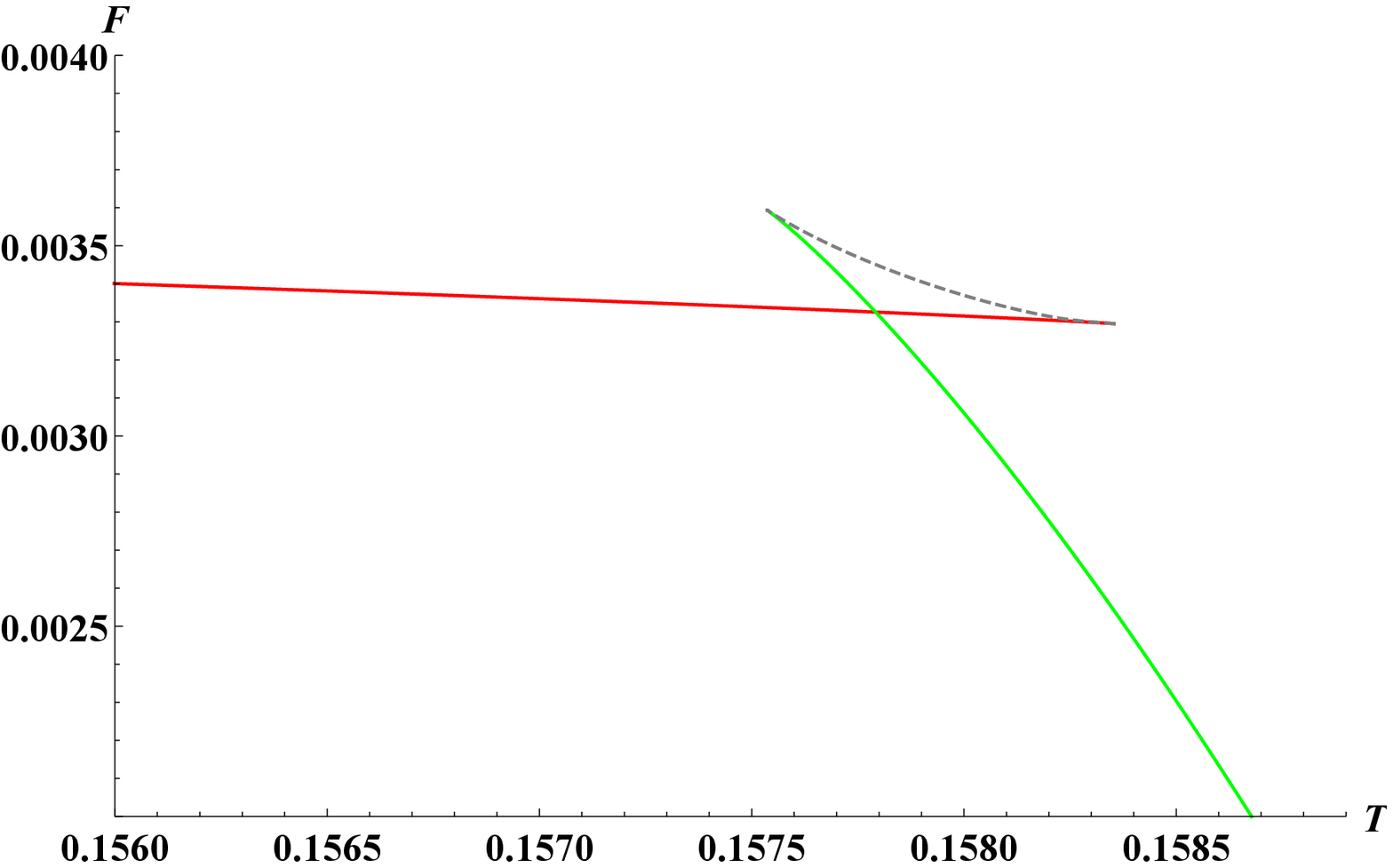}
        \label{C=44}
	\end{minipage}}
        \caption{The Gibbs free energy $F$ with temperature $T$ for the case $Q=0.043$. There are three critical points of central charge $(C_{\alpha},C_{\beta},C_{\gamma})$. Thus we fix the value of central charge as 34, 39.5, 42 and 44 for different cases. The thermodynamically stable branches are represented by the red/green/blue solid lines respectively. The unstable branches are described by the gray dashed lines.}
        \label{Q=0.043}
\end{figure}

For this case, we take $Q=0.043$, and three critical points are $(C_{\alpha},C_{\beta},C_{\gamma})=(38.7043,41.1913,43.0675)$. The diagrams of Gibbs free energy with temperature are plotted in Fig. (\ref{Q=0.043}).
In Fig. (\ref{C=34}) and Fig. (\ref{C=39.5}), the curves and phase structures are similar to Fig. (\ref{C=20}) and Fig. (\ref{C=30}) respectively. For $C_{\beta}<C<C_{\gamma}$, the new stable intermediate black holes branch and swallowtail appear. The position of the new swallowtail is different from that in Fig. (\ref{Q=0.035}), but both cases represent the same phase transition structures which is the first order phase transiiton of small/large black holes. When $C$ increases to $C_{\gamma}$, the new stable branch and second swallowtail disappear, leaving only two stable branches and one swallowtail on the diagram, which represent the phase transition of small/large black holes. 

\paragraph{$Q_3<Q$}
~{}

For this case, there is only one critical point. We set $C=0.08$, and the critical point is $C_c=40.2888$, 
\begin{figure}[H]
	\centering
	\includegraphics[scale=0.42]{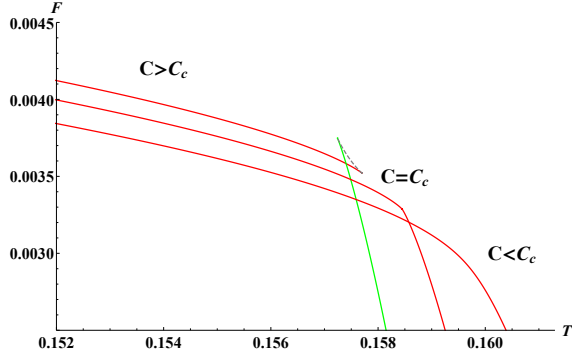}
	\caption{The Gibbs free energy $F$ with temperature $T$ for  $Q=0.08$. The red/blue lines represent the stable branches and grey dashed line indicates the unstable branch. There is only one critical point $C_c=40.2888$, and we take the value of central charge $C$ are 35, $C_c$ and 44 from bottom to top.}
	\label{Q=0.08}
\end{figure}
For the case of only one critical point, Fig. (\ref{Q=0.08}) shows that for $C\leqslant C_c$ the red solid line and blue dotted line represent the case of no phase transition behavior. For $C>C_c$, there are two stable branches represented by the red/green lines and one unstable intermediate black holes branch represented by the gray dashed line. The swallowtail shows the same first order phase transition structure of small/large black holes as our analysis before in Fig. (\ref{free energy fix the central charge C 4D}) and Fig. (\ref{f-T 5D}).
\section{Conclusion}
In this paper, we studied the thermodynamics of charged Gauss-Bonnet AdS black holes by regarding the cosmological constant $\Lambda$, Newton gravitational constant $G$ and Gauss-Bonnet coupling coupling $\alpha$ as variables. With this consideration, we introduced the central charge into the thermodynamics of black holes and obtained the mixed form of first law of thermodynamics in $D\geq 4$. The new form is mixed because it considers the contributions of variables in bulk and boundary field theory. Therefore, we can study the properties of critical central charge which depend on the Gauss-Bonnet term $\alpha$. We analysed the property of critical central charge for different electric charge $Q$ by numerical calculation. Then we discussed the phase transition structures based on different central charge $C$ and obtained the first order phase transition behaviors of small/large black holes in different dimensions. Besides, we find that there are three critical points in $D=6$ for some parameters $Q$, which implies more complex phase transition structures. A new stable phase  which repersents the intermediate black holes occurs. There exist second order phase transition structures of small/intermediate/large black holes, and the phase of three kinds of black holes can coexist at the triple point. The phase transition behavior of varivation of central charge $C$ in our work is similar to that of varivation of pressure $P$ for usual studies \cite{Wei:2014hba}. The phase transition  occurs when $C$ increases until it exceeds the critical value, while $P$ decreases to less than the critical value. Furthermore, it is of our interest to study thermodynamics in different phase spaces based on the modification of the first law, and we can extend the thermodynamics of charged Gauss-Bonnet AdS black holes in the context of the AdS/CFT correspondence.

\begin{acknowledgments}
We are grateful to Ningchen Bai and Aoyun He for useful discussions. This work is supported in part by NSFC (Grant No.11947408 and 11875196 and 12047573).
\end{acknowledgments}

\end{document}